\documentclass[journal,twocolumn]{IEEEtran}
\ifCLASSINFOpdf
  \usepackage[pdftex]{graphicx}
  \graphicspath{{../pdf/}{../jpeg/}}
  \DeclareGraphicsExtensions{.pdf,.jpeg,.png}
\else
  \usepackage[dvips]{graphicx}
  \graphicspath{{../eps/}}
  \DeclareGraphicsExtensions{.eps}
\fi
\usepackage{color}
\usepackage{epsfig}
\usepackage{booktabs}
\usepackage{epstopdf}
\usepackage{amssymb}
\usepackage{mathtools}
\usepackage{algorithmic}
\usepackage{array}
\usepackage{mdwmath}
\usepackage{mdwtab}
\usepackage{bm}
\usepackage{mathrsfs}
\usepackage{epsfig}
\usepackage{subfigure}
\usepackage{algorithmic}
\usepackage{amsmath}
\usepackage{mathrsfs}
\usepackage{cite}
\usepackage{url}
\usepackage{tikz}
\usepackage{pbox}
\usetikzlibrary{quantikz}

\DeclareGraphicsExtensions{pst,eps}
\graphicspath{{./fig/}}

\newcommand{\qI}{{\bf I}}
\newcommand{\qww}{{\bf w}}
\newcommand{\qp}{{\bf p}}

\newcommand{\qq}{{\bf q}}
\newcommand{\qh}{{\bf h}}

\newcommand{\qx}{{\bf x}}
\newcommand{\qK}{{\bf K}}

\graphicspath{{fig/}}
\newcommand{\be}{\begin{equation}} \newcommand{\ee}{\end{equation}}
\newcommand{\bea}{\begin{eqnarray}} \newcommand{\eea}{\end{eqnarray}}

\begin{document}

\title{Hybrid Quantum-Classical Neural Networks for Downlink Beamforming Optimization}
\author{Juping Zhang,
        Gan Zheng,~\IEEEmembership{Fellow,~IEEE,}
        Toshiaki Koike-Akino, ~\IEEEmembership{Senior Member, IEEE},
        Kai-Kit Wong,~\IEEEmembership{Fellow,~IEEE},
        and Fraser Burton
  \thanks{J. Zhang is with the Department of Electronic and Electrical Engineering, University College London, London, WC1E 6BT, UK (Email: juping.zhang,@ucl.ac.uk).}
     \thanks{G. Zheng is with the School of Engineering, University of Warwick, Coventry, CV4 7AL, UK (Email: gan.zheng@warwick.ac.uk).}
     \thanks{K. K. Wong is affiliated with the Department of Electronic and Electrical Engineering, University College London, Torrington Place, WC1E 7JE, United Kingdom and he is also affiliated with Yonsei Frontier Lab, Yonsei University, Seoul, Korea. (Email: kai-kit.wong@ucl.ac.uk).}
     \thanks{T. Koike-Akino is with Mitsubishi Electric Research Laboratories (MERL), Cambridge, MA 02139, USA (Email: koike@merl.com).}
     \thanks{F. Burton is with Applied Research,    British Telecom, Adastral Park, Martlesham Heath, Ipswich, IP5 3RE. (E-mail: fraser.burton@bt.com).}
} 
\maketitle
 
\begin{abstract}
This paper investigates quantum machine learning to optimize the beamforming in a multiuser multiple-input single-output downlink system.  We aim to combine the power of quantum neural networks and the success of classical deep neural networks to enhance the learning performance. Specifically, we propose two hybrid quantum-classical neural networks to maximize the sum rate of a downlink system. The first one
proposes a quantum neural network employing parameterized quantum circuits that follows a  classical convolutional neural network. The classical neural network can be jointly trained with the quantum neural network or pre-trained leading to a fine-tuning transfer learning method. The second one designs a quantum convolutional neural network to better extract  features followed by a classical deep neural network. Our results demonstrate the feasibility of the proposed hybrid neural networks, and  reveal that the first method can achieve similar sum rate performance compared to a benchmark classical neural network with significantly less training parameters; while the second method can achieve higher sum rate especially in presence of many users still with less training parameters.  The robustness of the proposed methods is verified using both software simulators and hardware emulators considering noisy intermediate-scale quantum devices.

\end{abstract}

\section{Introduction}
Multi-antenna technique or multiple-input and multiple-output (MIMO) has played an important role in the evolution of wireless communications.
Continued mobile traffic growth, without any associated increase in revenue as customers adopt unlimited data bundles, requires a focus on cost reduction and hence efficiency improvements, both for energy and spectrum. MIMO beamforming is a popular transmit strategy to deliver improvements in both power efficiency and spectral efficiency. Optimization is thus critical to achieve high performance of MIMO systems that employs beamforming. In the early literature, the optimization of beamforming largely rely on numerical algorithms according to different design objectives such as minimization of total transmit power \cite{rashid1998transmit,bengtsson1999optimal}, maximization of  signal-to-interference-plus-noise ratio (SINR) \cite{schubert2004solution} and sum rate  \cite{shi2011iteratively,christensen2008weighted}. The challenge with numerical algorithms is that they are iterative in nature, often experiences  low convergence and cannot scale well as the system becomes large. Consequently, these algorithms cannot meet  the increasing demand of high data rate, high reliability and ultra low latency in the beyond  fifth-generation (5G) and the sixth-generation (6G) mobile communications systems. Improvement of MIMO beamforming algorithms requires real time access to data measured from the radio channel between user and base station (BS). The ultimate capacity of the radio interface with massive MIMO antennas is significantly greater than being achieved today. Users must move slowly enough for uplink channel information to be an accurate estimate for the downlink channel, so a rapidly converging optimization algorithm can achieve the beamforming gain for more rapidly moving users. With greater knowledge of the customer service and user distribution and with the use of rapid machine learning  to improve MIMO  algorithms, there is an opportunity to significantly improve 5G and 6G spectral efficiency. 

To speed up the optimization algorithms, deep learning based techniques have been exploited to approximate the near-optimal beamforming solution. The main idea is to learn the mapping from the input channel state information to the output beamforming solution by training a deep neural network without the complex and iterative numerical procedure. This can be done offline infrequently, and when used in real time, the trained model can predict the beamforming directly. This idea has been successfully used to develop a series of deep learning algorithms using the fully connected deep neural networks\cite{huang2018unsupervised,huang2019fast}, convolutional neural networks (CNN) \cite{xia2019deep,dl_sumrate} and graph neural networks (GNN) \cite{GNN-zhang,GNN-yang},  to address different beamforming optimization problems. However, even with deep neural networks, realizing the benefits of optimized beamforming is challenging as classical neural networks cannot keep pace with the rapidly increasing user demand in 6G communications and the computational-heavy optimization will become a bottleneck again.  Take beam selection in multicell massive MIMO systems for example: when there are hundreds of beams, one needs to find the best solution from billions of beam combinations and its associated complexity cannot be afforded by using classical algorithms \cite{beamselection}.  If  MIMO techniques continue to thrive in 6G communications, there must be more efficient optimization methods to design the beamforming strategy.

Recently researchers have started to explore quantum computing as a promising alternative technology to broadly address the resource allocation in wireless communications systems.  Unlike the bit in classical computing which can either take the state 0 or 1,  in quantum computing, a qubit is  the basis unit and  can have the states of $\ket{0}$ or $\ket{1}$ (which will be explained in Appendix), or be in the superposition of the two states simultaneously. In general, by using $b$ bits in classical computing, only one of the $2^b$ possible combinations can be represented; while with $b$ qubits,   a superposition of all $2^b$ states can be represented and this parallel processing shows significant  improvement over classical computing.  Quantum technology is fast developing and the progression of the number of qubits has been rapid over the years. Starting from 2 qubits by IBM, MIT and UC Berkeley in 1997, IBM achieved 5 qubits in 2016 and increased it to 50 in 2017. Google unveiled its 72-qubit chip in 2018. IBM then released 433-qubit and 1,121-qubit processors in 2022 and 2023, respectively, and set out a key milestone of achieving a 100,000-qubit system by 2033.

It is the superposition and entanglement that make quantum computing so powerful and be able to accomplish tasks intractable for classical computing. For instance, Shor's algorithm can factor a large integer number in polynomial time which is widely
believed to have no efficient solution on a classical computer and it has the potential to break  public-key cryptography schemes \cite{Nielsen-book}. For instance, the Grover's algorithm \cite{Nielsen-book}, one quantum search algorithm, can find a desired value with high probability in a database with $N$ entries using just $O({\sqrt{N}})$ evaluations of the function while classical computing requires at least  $O(N)$ evaluations.

 Quantum search algorithms have been proposed to improve the performance of wireless communications systems such as multiuser detection \cite{quantum-MUD}, joint channel estimation and data detection \cite{quantum-joint-channel-data}, vector-perturbation based precoding \cite{quantum-precoding}, and routing in multi-hop communications  \cite{quantum-routing}, and the performance improvement has been verified via simulations.   This is termed as quantum-assisted communications using quantum search algorithms, and a comprehensive survey can be found in \cite{quantum-search}.
Quantum annealing is another popular algorithm that has been used to solve discrete  problems as long as the objective function formulations are binary quadratic models. Examples solved by quantum annealing include  optimization of vector  precoding \cite{Annealing-MIMO} and phase optimization of reflective metasurfaces \cite{Annealing-RIS} on an advanced QA hardware, the D-Wave 2000-qubit (DW2Q) quantum adiabatic optimizer machine \cite{dwave}.  A hybrid quantum-classical algorithm called quantum approximate optimization algorithm was recently proposed to solve NP-hard problems \cite{QAOA}. This algorithm was successfully applied to solve the channel decoding problem and validated on a real IBM quantum computer \cite{QAOA-decoding}.

 Quantum machine learning (QML) is one of the most active research areas that combines the advantages of quantum computing and machine learning \cite{qml}. By taking advantage of quantum effects such as entanglement and superposition, QML can process data more efficiently and thus achieve faster convergence rate and    increased accuracy in prediction which often leads to improved end performance.
 It is proved in \cite{expressive-power} that multiple-layer parametrized quantum circuits (PQCs), which is one form of the quantum neural network (QNN), has stronger expressive power than a classical neural network when approximating the distributions generated by instantaneous quantum polynomial time (IQP) circuits. This is because those distributions cannot be sampled efficiently by any classical neural network, which partially explains the reason of  ``quantum supremacy''.
A separate study \cite{QNN-power} adopts the metric of effective dimension to explain the supremacy of QNN. Effective dimension is to estimate the size that a model occupies in model space – the space of all possible functions for a particular model class, where the Fisher information matrix serves as the metric.  Intuitively, the higher the effective dimension, the better the performance of neural networks. For instance, classical neural networks exist in very high-dimensional parameter spaces, but their true size, represented by effective dimension, is typically far smaller \cite{QNN-IBM}. It is shown in \cite{QNN-power}  through numerical experiments  that QNNs are able to achieve significantly higher effective dimensions than their classical counterparts.  Quantum generative adversarial learning networks have been introduced that may exhibit an
exponential advantage over classical adversarial networks  when the data consist of samples of measurements made on high-dimensional spaces \cite{generative}.
Despite QML being an emerging research area, it has already found wide applications in many information  and engineering related areas recently, including speech recognition \cite{Recognition1}\cite{Recognition2}, anomaly detection \cite{Anomaly1}\cite{Anomaly2},   earth observation and remote sensing \cite{RS1} \cite{RS2}, and wireless communications.
A comprehensive review on  quantum machine learning  and its applications to  resource allocation and  network security for 6G wireless networks can be found in \cite{Trung-qml}.  A QNN and a reinforcement-learning-inspired QNN are presented in \cite{Narottama-RA} to reduce the time complexity of resource allocation in  non-orthogonal multiple access systems while maintaining the performance. This method is extended to the application of transmitter-user assignment in the cell-free MIMO scenario \cite{Narottama-MIMO}. A QNN is developed in \cite{Ni-EE} to improve the energy efficiency and it exhibits a slightly faster convergence speed than its classical counterpart. In \cite{Green}, quantum reinforcement leaning (QRL) is designed to jointly optimize relay and transmit power selection and its advantage is shown compared to state-of-the-art techniques in terms of convergence speed and network utility. A novel quantum-inspired experience replay (QiER) framework is proposed in \cite{Li-QDRL} to help a UAV find the optimal flying direction and thus its trajectory towards the destination. It is shown to  achieve more efficient learning, require  less hyper-parameters and minimum time cost compared to classical deep reinforcement learning.

While the above initial studies mostly assume that the quantum machines operate in a perfect and desired way,
one of the major challenges of quantum computing in the noisy intermediate-scale quantum (NISQ) era  is to deal with errors due to limited resources
available \cite{nisq,ibm-noise}. Quantum systems are extremely vulnerable to environmental effects which rapidly decohere their quantum
properties. While eventually quantum error correction techniques may have to be used  to ensure fault-tolerant quantum computation, they are not practical to implement currently due to high overhead in terms the number of qubits and quantum gates. Therefore, we have to assume that NISQ devices cannot mainly rely on quantum error correction and other more robust utilization of noisy quantum devices must be sought. In this regard,
variational quantum algorithms (VQA) \cite{Parameterized,VQA} have been proposed as a promising method to achieve quantum advantages on the NISQ devices. VQA has a common hybrid structure including a parameterized quantum circuit with its measured expectation output contributing to the computing of the cost function.  A classical processor will optimize the circuit parameters by minimizing the cost function. This gives rise to the hybrid quantum-classical nature that combines the power of quantum computing with the success of existing classical neural networks \cite{transfer}. Because the circuit has relatively shallow depth, the overall algorithm is  resilient against quantum noise caused by  decoherence and imperfect gate circuits as studied in Section IV.  The application of VQA-based hybrid machine learning in wireless communications is still rare. In \cite{Toshi-WiFi-Sensing}, variational quantum circuit are used to build QNN  as a new QML approach for Wi-Fi sensing. The proposed transfer learning framework with a  small-scale QNN can achieve greater than 90\% accuracy, comparable to a large-scale DNN, and can improve the robustness against domain shifts across Wi-Fi scanning sessions. A method to automate the design of quantum circuits (also known as ansatz) is first proposed in \cite{Toshi-ISC} and applied in the scenario of WiFi integrated sensing and communications. It is shown that a small-scale QNN can achieve state-of-the-art performance comparable to a large-scale DNN, in human pose recognition.

Motivated by the aforementioned efforts and lack of variational quantum algorithms to resource optimization in wireless communications, this paper investigates hybrid quantum-classical neural networks employing variational quantum circuits to  optimize multiuser beamforming with the design objective of maximizing the sum rate.  We refer ``hybrid neural networks'' as a hybrid use of both classical neural networks and quantum neural networks jointly. A beamforming optimization that can achieve a higher sum-rate in a shorter time than a classical algorithm can equate to more efficient network resource utilization serving more users.
 To the best of our knowledge, this paper is the first study that introduces hybrid quantum-classical neural networks to solve challenging resource optimization problems in wireless communications and demonstrates their advantages over classical neural networks.
Our main contributions are summarized as follows:
\begin{itemize}
\item We introduce the framework of hybrid quantum-classical neural networks to optimize the beamforming of multiuser multi-antenna systems with the aim to enhance the capability of classical neural networks to maximize the sum rate. It synthesises benefits of the well studied classical deep learning neural networks, the quantum parallelism and quantum entanglement  that allow more efficient and effective learning,  and shallow quantum circuits that are robust against noise.

\item We propose two specific hybrid quantum-classical neural network structures to achieve the above aim. {{The first one employs a small QNN  to replace large fully connected layers to exploit the extracted features by a classical CNN. Thanks to the small-scale QNN,   the amount of trainable parameters  increases only linearly with the number of users and the number of antennas at the BS. This is in stark contrast to the benchmark classical neural network approach in which the amount of trainable parameters increases linearly with the product of the number of users and the number of antennas at the BS.}} The second one investigates quantum CNN to better extract the features from data which will be fed into the classical fully connected layers to achieve superior performance.

\item We carry out extensive simulations to evaluate the learning performance and the sum rate  of the proposed hybrid structures and demonstrate their advantages in terms of the improved sum rate and reduced trainable parameters. Moreover, we assess their performance in noisy devices using both software simulations and hardware emulators, and confirm their resilience against errors.
\end{itemize}

 Note that in this paper we employ a  universal   general-purpose  gate-based quantum computing platform which   can be applied to a wide  range of applications and is not solely dedicated for this specific beamforming optimization problem in wireless communications.

The remainder of this paper is organized as follows. Section \ref{system_model} introduces the system model, problem formulation and classical neural network design to solve the sum rate maximization problem. Section \ref{algo} provides details of the proposed hybrid learning framework and two hybrid neural network structures with analysis on the comparison of trainable parameters with the classical counterpart.  Simulation results and conclusions are presented in Section \ref{simu} and Section \ref{conc}, respectively. {{We leave the fundamental concepts of quantum computing that are relevant to the development of our propose algorithms to the Appendix  for readers' convenience.}} 

{\em Notions:} All boldface letters indicate vectors (lower case) or matrices (upper case).  The superscripts   $(\cdot)^H$ and  $(\cdot)^{-1}$ denote the conjugate transpose and the inverse of a matrix, respectively. In addition, $\|\mathbf{z}\|_2$  denotes  the $L_2$  norm of a complex vector $\mathbf{z}$.  The operator $\mathcal{CN}(0, \mathbf{\Theta})$ represents a complex Gaussian vector with zero-mean and covariance matrix $\mathbf{\Theta}$. $\mathbf{I}_M$ denotes an identity matrix of size $M\times M$.  

\section{System Model, Problem Formulation and Classical Neural Network}\label{system_model}
\subsection{System Model and Problem Formulation}
We consider a multi-input   single-output (MISO) downlink  system where  a BS with $N_t$ antennas serves $K$ single-antenna users. The received signal at user $k$ can be written as
{\begin{align}
\mathbf{y}_k = \mathbf{h}_k^H\mathbf{w}_k\mathbf{s}_k+n_k,
\end{align}} where $\mathbf{h}_k\in\mathbb{C}^{N_t\times1}$ denotes the channel vector between the BS and user $k$, $\mathbf{w}_k$ and $\mathbf{s}_k\sim\mathcal{CN}(0,1)$ denote the transmit beamforming vector and the information-bearing   signal for user $k$ with normalized power, respectively. The additive Gaussian white noise (AWGN) is given by $n_k\sim\mathcal{CN}(0,\sigma^2)$. As a result, the received SINR   at user $k$ is expressed as
{ \begin{align}\scriptsize
\gamma_k = \frac{|\mathbf{h}_k^H\mathbf{w}_k|^2}{\sum_{j\neq k}^K|\mathbf{h}_k^H\mathbf{w}_j|^2+\sigma^2}, \forall k.
\end{align}}
Based on the aforementioned system setup,  we consider the sum rate maximization problem   under a total power constraint $P$  which is formulated as
{ \begin{align}\label{generalproblem}\scriptsize
\max_{\mathbf{w}_k, k=1,\ldots,K}~      \sum^K_{k=1} \log_2(1+\gamma_k), ~~~~\mathrm{s.t.}~~\sum_{k=1}^K\|\mathbf{w}_k\|_2^2\leq P,
\end{align}}
where we have assumed that perfect channel state information is available. Normally the optimization of beamforming is nonconvex and difficult to solve. Specifically, the sum rate optimization  \eqref{generalproblem} a  well-known challenging problem and there is no practical algorithm to find its optimal solution. Instead, it is often solved in an iterative way such as using the weighted minimum mean squared error (WMMSE) algorithm
 \cite{shi2011iteratively,christensen2008weighted}. Although the WMMSE algorithm achieves close to optimal solutions with a large number of iterations, its iterative nature brings high complexity and therefore it is not suitable for practical implementation in particular for low-latency systems.

\subsection{Overview of A Classical Deep Neural Network Solution}\label{classical}
Deep learning has emerged as a new paradigm to learn the mapping from the channel state information to the beamforming solution offline, and then infer the near-optimal solution in real time with low complexity and computational latency \cite{huang2018unsupervised,huang2019fast,xia2019deep}. {{Learning the beamforming solution directly is a nontrivial task because of the high-dimensional beamforming vectors which contain $2N_tK$ real variables. To facilitate the deep neural network design, \cite{xia2019deep} has proposed a model-based approach that aims to first  learn the key features of the problem  \eqref{generalproblem} with a reduced dimension, and then the beamformig solution can be recovered from the key features and channel state information.}}

 To be specific,  the optimal parameterized downlink beamforming vectors  can be expressed as \cite{bjornson2014optimal}:
\begin{equation}\label{solution struc of sumrate}
  \qww_k^{*}=\sqrt{p_k}\frac{(\qI_N+\sum^K_{k=1}{\frac{q_k}{\sigma^2}\qh_k\qh_k^H})^{-1}\qh_k}{||(\qI_N+\sum^K_{k=1}{\frac{q_k}{\sigma^2}\qh_k\qh_k^H})^{-1}\qh_k||_2}, \forall k,
\end{equation}
 where  $\{\qp\}$ is the downlink power vector with a sum of $P$ and $p_k = \|\qww_k\|^2$. $\{\qq\}$ denotes the virtual uplink power for a symmetric scenario where the downlink and the virtual uplink channels are equally strong and have well separated directivity, while both channels have the same SINR region for all users and the same total transmit power.
$\{\qq\}$ and $\{\qp\}$ will be exploited as the low-dimensional features. In other words, instead of predicting the beamforming vectors directly, the neural network will
predict the power vectors $\{\qq\}$ and  $\{\qp\}$ and then with the available channel information, the required beamforming solution can be recovered by using \eqref{solution struc of sumrate}. We can see that by doing so, the output dimension of the neural network is reduced from $2N_tK$ to $2K$, which leads to much reduced  complexity and improved accuracy.

 Because the problem is \eqref{generalproblem} is nonconvex and it is difficult to generate enough labelled data, we adopt unsupervised learning  to train the neural network. The resulting loss function is thus the negative average  sum rate directly, i.e.,
\begin{equation}\label{loss2}
  \text{Loss}=-\frac{1}{2KN}\sum^N_{n=1}\sum^K_{k=1} \log_2\left(1+\gamma^{(n)}_k\right),
\end{equation}
where $N$ is the batch size. {{The beamforming vectors need to meet the total transmit power constraint in \eqref{generalproblem}, while this may not be satisfied by the output of the neural network. In the construction of the above loss function, before we calculate the SINR, we normalize   the output of the neural network such that the total transmit power is always $P$.}}

\begin{figure}[t]
\centering
\includegraphics[width=3.5in]{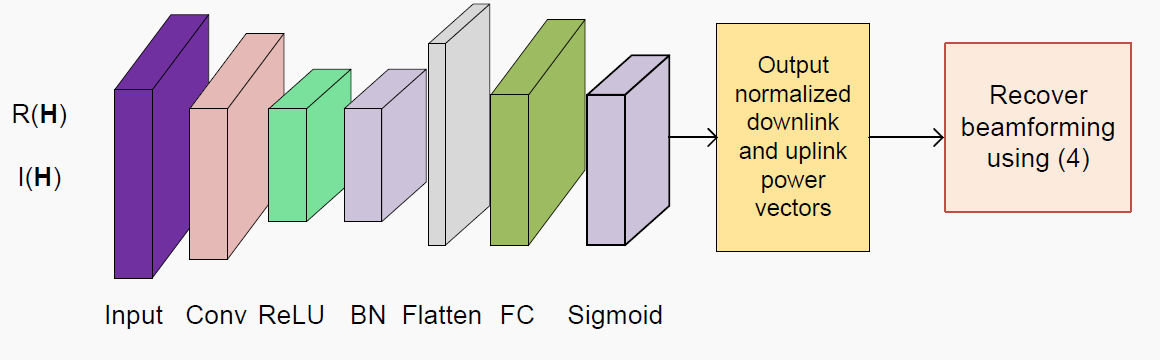}
\caption{The classical neural network structure to solve \eqref{generalproblem}.}
\label{fig:classical_NN}
\end{figure}

The network structure is adapted from the model-based beamforming neural network framework proposed in \cite{xia2019deep} to learn the power vectors, as illustrated in Fig. \ref{fig:classical_NN}.
 This framework is composed of a  CNN  architecture, one batch normalization (BN) layer, one flatten layer, one fully connected  (FC) layer and one Sigmoid activation layer at the end. The output includes the uplink and downlink power vectors and they will be used to find the original beamforming vectors using \eqref{solution struc of sumrate}. The CNN layer is useful to extract features and it  applies $F$ kernels of size $m\times m$, one stride, and one padding. The complex channel input is split into two real value inputs, so the input dimension of the input layer is $K\times 2N_t$.  {{The Adam optimizer is adopted \cite{kingma2014adam} for the optimization of the neural network with a learning rate of 0.001. Detailed description of the classical neural network structure is provided in Table \ref{BNN para}.}}
\begin{table}[tbp]
\centering
\caption{ Parameters of the neural network modules of the classical neural network structure in Fig. 1.}
\setlength{\tabcolsep}{1.5mm}{\begin{tabular}{ll}
\hline
Layer& Parameter\\
\hline
Layer 1 Input& \begin{minipage}{4cm} \vspace{0.1cm}Input of size $K\times 2N_t \times  1$ \vspace{0.1cm}\end{minipage} \\
Layer 2 2D convolutional&  \begin{minipage}{4cm} \vspace{0.1cm} $F$ kernels of size $m\times m$, zero padding  1, stride 1\vspace{0.1cm}\end{minipage} \\
Layer 3 Activation& ReLU\\
Layer 4 Batch normalization&   non-trainable parameters: momentum=0.1, $\epsilon=10^{-5}$\\
Layer 5 Flatten \\
Layer 6 Fully-connected)& $2K$ neurons\\
Layer 7 Output Activation  & Sigmoid\\
\hline
\end{tabular}}
\label{BNN para}
\end{table}

\section{Hybrid Quantum-classical Neural Networks for Beamforming Optimization}\label{algo}
The aim of this paper is to leverage the latest quantum technology to enhance the classical neural networks to optimize the downlink  beamforming design. In this section, we propose a framework of hybrid quantum-classical neural network that introduces quantum layers into the classical neural network presented in Section II (illustrated in Fig. \ref{fig:classical_NN}) to improve its performance.

A  QNN  layer that can be incorporated in the classical neural network usually contains the following three components:
\begin{itemize}
  \item Data embedding. It  transforms classical information into  quantum states in higher dimensional Hilbert space, and is an essential first step in using quantum machine learning algorithms to solve classical problems.
      It can be seen as a quantum circuit composed of quantum gates to  a set of $\ket{0}$ quantum nodes and the parameters of the gates are determined by the classical data. How to design efficient data embedding is still an active research area, and there are widely used data embedding methods. Below we introduce two of them \cite{baidu} that will be used in our proposed hybrid neural network.
      \begin{enumerate}
        \item Angle encoding.
        Angle encoding makes use of rotation gates to encode classical information $\{{x_i}\}$. The classical information determines angles of rotation gates:
        \be
            \ket{\psi} = \otimes_{i}^n R(x_i) \ket{0^n},
        \ee
        where $R(\cdot)$ denotes a rotation gate about the X, Y or Z-axis of the Bloch sphere and in this paper we use $R_y$ introduced in the Appendex. The number of qubits required for encoding is the same as  the length of the classical vector. For instance, suppose the classical data is $\qx = [\pi ~~ \pi ~~ \pi]$, then using $R_y$ gate the corresponding quantum state will be $\ket{111}$.

        \item Amplitude encoding. This method encodes a normalized classical vector $\qx$  of length $N$ into amplitudes of an $n$-qubit quantum state:
         \be
            \ket{\psi} = \sum_{i=1}^N x_i \ket{i},
         \ee
         where $N=2^n$, $x_i$ is the $i-$th element of the vector $\qx$ and $\{\ket{i}\}$ is the computational basis for the Hilbert space.
         For instance, for a classical vector $\qx = [\frac{1}{2} ~~\frac{1}{2}~~-\frac{1}{2}~~-\frac{1}{2}]^T$, the amplitude encoded quantum state can be expressed as
         \be
            \ket{\psi} = \frac{1}{2} \ket{00} + \frac{1}{2}\ket{01}- \frac{1}{2}\ket{10} - \frac{1}{2}\ket{11}.
         \ee
         As a $n$-qubit system can provide $2^n$ amplitudes, amplitude encoding only requires $n= \lceil\log_2(N)\rceil$ qubits, compared to $N$ qubits for the angle embedding.

      \end{enumerate}

  \item Parameterized quantum circuit.  A parameterized quantum circuit is the main circuit following the data embedding. Usually one layer of circuit includes entangling operations such as CNOT and parameterized  rotations using single-qubit quantum gates. The rotation angles of the rotation gates are trainable parameters similar to the weights of classical neural networks. The whole circuit can have multiple identical or different layers to improve its expressive power.

  \item Measurement. The outputs of the quantum states are retrieved by measuring the qubits' states in the computational basis (usually the Pauli-Z basis). This process essentially converts quantum states back to classical data which will then be used as input for the rest classical neural network layers.

\end{itemize}
It is worth mentioning that the initialization for the qubits is done by setting their states to $\ket{0}$ and the gates   parameters are initialized by using a uniform distribution in $[-2\pi,  2\pi]$. 

There are different ways in which QNN can be incorporated into classical neural networks. It can either make better use of features extracted by the classical layers or can more effectively extract features to be used by classical layers. In this paper, we propose  two hybrid structures that will be presented below.

\subsection{A Hybrid QNN Structure}
\begin{figure}[h]
\centering
\includegraphics[width=3.5in]{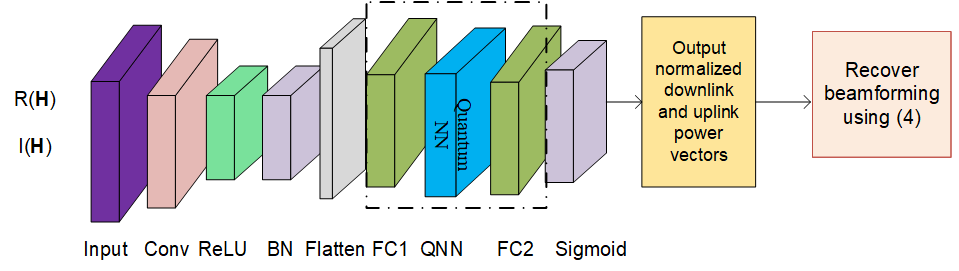}
\caption{The proposed hybrid   QNN structure to solve \eqref{generalproblem}.}
\label{fig:CNN_QNN}
\end{figure}

The first hybrid neural network structure is to add a QNN after the classical CNN as illustrated in Fig. \ref{fig:CNN_QNN}. Compared to the classical neural network in Fig. \ref{fig:classical_NN}, this structure replaces the fully connected layer 'FC' with a QNN layer together with two smaller fully connected layers 'FC1' and 'FC2'. The CNN is well known to extract features effectively, and the motivation of this design is to leverage QNN to better exploit the features extracted by the CNN than the original fully connected layer.
The details of the QNN  is shown in Fig. \ref{fig:CNN_QNN:circuit}. {{As described in the previous subsection, the proposed QNN has the embedding layer, parameterized quantum circuit layer and measurement layer with details below.}}

{{1) Data embedding layer. The data embedding layer employs angle encoding to convert the classical data to quantum states. \footnote{The encoding methods for QNN and QCNN are chosen based on the empirical results in Section V.} Suppose the output length of the flatten layer is $O$ which will become clear in Section III. C. Then the first small fully connected layer (`FC1' in Fig. \ref{fig:CNN_QNN}) with the activation function `tanh' has $Q$ neurons and generates $Q$ classical input data to the QNN, denoted as $\{c_i\}_{i=1}^Q$. For angle encoding, we first apply the Hadamard gate and then the $R_y$ gate to encode classical channel data $\{c_i\}$ to quantum data, i.e.,
         \be
            \ket{\psi} = \otimes_{i}^Q R_y(c_i) H \ket{0^Q}, 
        \ee
and we can clearly see that the classical bits  determine   the angle of   $R_y$ gates. This process is illustrated in the first part of Fig. \ref{fig:CNN_QNN:circuit}.\\
$~~~~$2) Parameterized quantum circuit layer. After quantum data embedding,  there are $L$ parameterized circuits with an identical structure and the $l-$th circuit is shown in the middle of Fig. \ref{fig:CNN_QNN:circuit}. The qubits are first entangled through the CNOT gates and specifically, the $i-$th qubit is the control qubit and the $(i+1)-$qubit is the target qubit.  It is then followed by $R_y$ gates with rotation angles as trainable parameters, which are similar to the trainable parameters including weight and bias in classical neural network. All trainable parameters will be jointly optimized by a classical optimizer during the training process.  This multi-layer circuit structure  allows to increase the entanglement\footnote{According to \cite{Entanglement}, entanglement is necessary in a quantum algorithm on pure states if the algorithm is to offer an exponential speed-up over classical computation, although it alone may not be an essential resource for quantum-computational power.} of qubits and exploitation of the quantum space.

Note that the trainable parameters are different across different circuits, so the total number of trainable parameters is $LQ$. The input and output dimensions of the  parameterized quantum circuit layer are both $Q$.\\
$~~~~$3) The measurement layer. As illustrated in the rightmost part of Fig. \ref{fig:CNN_QNN:circuit}, after the  parameterized quantum circuits, the measurement layer converts the quantum states to the classical data. Specifically, it takes the expectation values of a number of measurements in  the Z-basis which projects the quantum state onto one of the states  $\ket{0}$ or $\ket{1}$, i.e., the eigenstates of the Pauli Z matrix. In our simulation, 1,000 circuit evaluations\footnote{The delay caused by the high number of circuit evaluations will exceed the 5G/6G latency budget. This will be improved as quantum technology advances and can be reduced by using less circuit evaluations at the cost of lower measurement quality.} are used to  estimate the expectation values. The output dimension of the measurement layer is still $Q$.

{{The following fully connected layer (`FC2' in Fig. \ref{fig:CNN_QNN}) has $2K$ neurons which takes the $Q$ dimensional output of the QNN as input.}} The whole hybrid neural network can be trained so that  the weights of the classical neural network and the rotation angles (i.e., $LQ$ in total) in the QNN can be jointly optimized so that the search space is expanded to utilize the higher expressive power of the quantum layer.  Apart from qubit entanglement, the main advantage of this method is that it requires less trainable parameters than the classical neural networks  since each layer of quantum circuit only has $Q$ parameters to optimize.  A detailed comparison  of neural network parameters is provided in Section IV.C.

 In addition, we can also exploit transfer learning in which the CNN module is pre-trained in the  classical deep neural network as a feature extractor and we keep its weights constant, so we only need to train the weights and rotation angles in the newly added fully connected layers and QNN. This method will further reduce the number of required trainable parameters and can be trained with fewer epochs.
\begin{figure*}
\centering
\begin{quantikz}
\lstick{$\ket{\psi_1}$}&\gate{H}\gategroup[4,steps=2]{{\sc Embedding}}& \gate{R_y(c_1)}   &\qw& \cdots & \ctrl{1}
\gategroup[4,steps=4,style={dashed,rounded corners,fill=blue!20, inner xsep=2pt}, background]{{\sc The $l$-th layer circuit}}
 & \gate{R_y(\theta_{l,1})}   & \qw & \qw  & \qw & \cdots    &\meter{}\gategroup[4,steps=1]{{\sc Measurement}}\\
\lstick{$\ket{\psi_2}$}&\gate{H}&  \gate{R_y(c_2)} &\qw& \cdots & \targ{}& \ctrl{1} & \gate{R_y(\theta_{l,2})}  & \qw  &\qw & \cdots &\meter{}\\
\lstick{$\ket{\psi_3}$}&\gate{H}& \gate{R_y(c_3)} &\qw& \cdots &  \qw &   \targ{}   &  \ctrl{1}   & \gate{R_y(\theta_{l,3})}   & \qw    & \cdots &\meter{} \\
\lstick{$\ket{\psi_4}$}&\gate{H}&  \gate{R_y(c_4)} &\qw& \cdots &\qw   & \qw &  \targ{}& \gate{R_y(\theta_{l,4})}  & \qw   & \cdots  &\meter{}
\end{quantikz}
\caption{The parameterized quantum circuit used in QNN, $Q=4$. $c_i$ is the $i$-th classical input data. There are $L$ entanglement layers in the middle part.}
\label{fig:CNN_QNN:circuit}
\end{figure*}
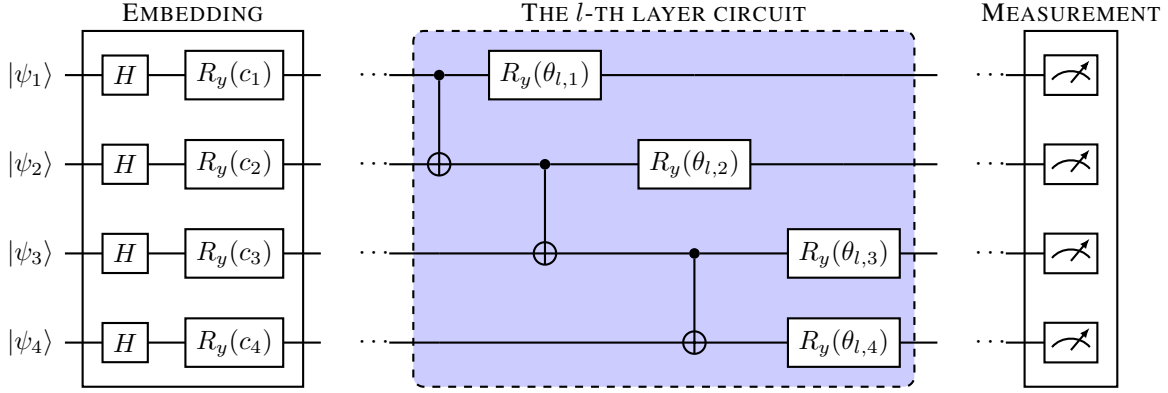

\subsection{A Hybrid Quantum CNN Structure}
 In the second hybrid scheme, we propose to enhance the classical neural network by introducing an additional quantum CNN (QCNN) layer before the CNN layer, as illustrated in Fig. \ref{fig:QCNN}. This is inspired by the quanvolution neural network originally proposed in \cite{QCNN}. CNN is well known for its capability to capture the spatial and temporal dependencies in the input data by applying  convolutional filters to local  subsections of the input to produce useful features. Another benefit of CNN is that it requires  many fewer parameters than fully connect layers via weight sharing. A QCNN layer operates in a similar way as CNN by extracting useful features through transforming local data and thus retains the advantages of CNN. The main difference is that a  quantum filter extracts features using purpose-designed parametrized quantum circuits, and possible performance gains come  from the fact that   QCNN can exploit a large quantum state space through controllable entanglement and interference \cite{nature-qml}. This would lead to improved machine learning models for  classical tasks.  

\begin{figure}[h]
\centering
 \includegraphics[width=3.8in]{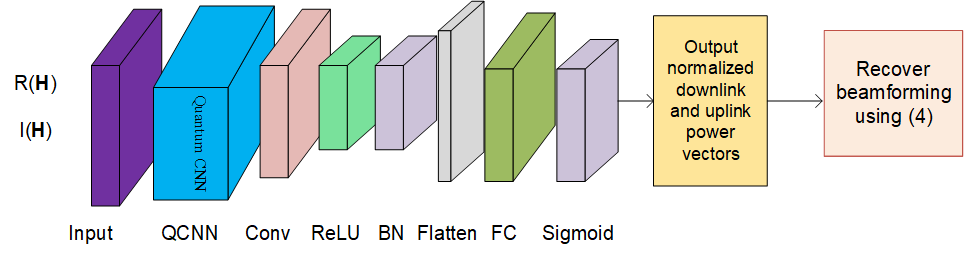}
\caption{The hybrid  QCNN  structure to solve \eqref{generalproblem}.}
\label{fig:QCNN}
\end{figure}

An example parameterized quantum circuit in QCNN is shown in Fig. \ref{fig:QCNN:circuit}. {{Similar to the QNN, it has the embedding layer, parameterized quantum circuit layer and measurement layer which are detailed below.\\
$~~~~$1) Data embedding.   We apply the parameterized quantum circuit to a small $2\times 2$ square of  input channel vector  with unit norm which contains four classical input (channel) data $c_1, c_2, c_3, c_4$. As shown in the first part of Fig. \ref{fig:QCNN:circuit}, we employ amplitude embedding to encode the classical data into  the following quantum states which   requires only two ($Q=\log_2(4)$) qubits:
        \be
            \ket{\psi} = \sum_{i=1}^4 c_i \ket{i},
         \ee
         where $\{\ket{i}\}$ belongs to the computational basis of two qubits, i.e., $\{\ket{00},\ket{01},\ket{10},\ket{11}\}$.\\
$~~~~$2) Parameterized quantum circuit layer.  After data embedding,  the classical data is represented by
two qubits  which are then followed by $L$  parameterized  circuits which has the same structure as that in the QNN method, i.e., it uses the CNOT  and $R_y(\theta)$ gates to generate qubits entanglement, and the angles of the $R_y$ gates are trainable parameters to be optimized. There are in total $2L$ trainable parameters.\\
$~~~~$3) Measurement layer. Finally, the two-qubit quantum states are measured to produce classical output using the same method as QNN by  taking the expectation values of  measurements in  the Z-basis.  As the QNN, we use  1,000 circuit evaluations to estimate the expectation values. Similar to CNN, by applying the parameterized quantum circuits to the whole input channel data (a $2\times2$ block each time), we can obtain the output of QCNN.
}}

The measurement output will be fed into the original  CNN layer as shown in Fig. \ref{fig:QCNN}. This structure can be generalized to $Q>2$ with $2^Q$ input data (e.g., a block of $2^{Q-1}\times 2^{Q-1}$) and is particularly attractive for NISQ devices since it only requires a small number of qubits and is thus robust against noise  which will be verified in Section IV.

Note that we use trainable QCNN unlike the original work in \cite{QCNN}. In \cite{QCNN}, QCNN is used as a data pre-processing mechanism to extract features and a random quantum circuit is adopted, so the rotation angles are not trainable. The quantum processed data is then used as the input to classical neural networks. This method limits the capability of QCNN while   our proposed solution  better exploits the quantum space by optimizing the rotation angles together with the weights of classical neural networks.

\begin{figure}[h]
\centering
\includegraphics[width=3.9in]{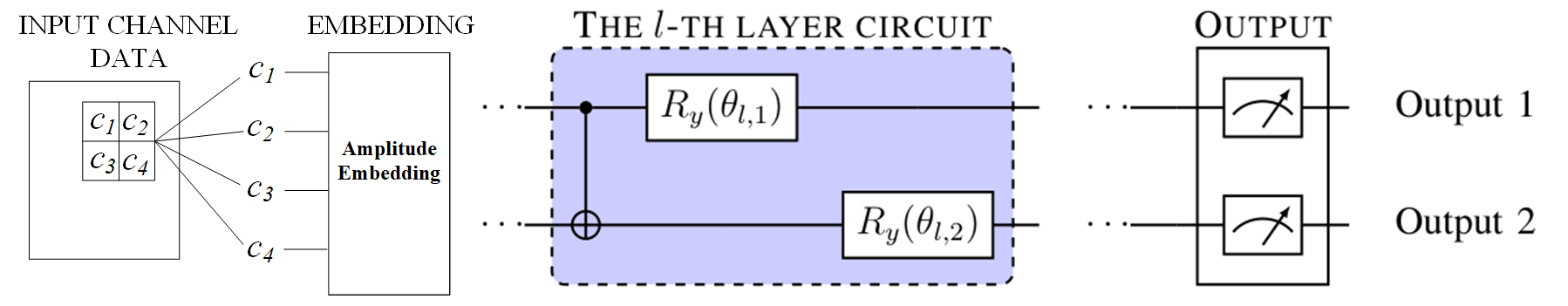}
\caption{Detailed circuit of the quantum convolutional layer. Here we assume it processes a block of $2\times2$ classical data, and the number of qubits is $Q=2$. $c_i$ is classical channel data. }
\label{fig:QCNN:circuit}
\end{figure}

\subsection{Parameter and Complexity Analysis}
{{A major advantage of the hybrid quantum-classical neural network is that it requires much less trainable parameters. This imlies that a hybrid quantum-classical neural network is more memory-efficient in training and storing the model and more efficient when deployed on quantum computers and it has better scalability. Therefore in this subsection, we analyze the amount of {{trainable parameters}} of our proposed hybrid neural networks and compare them with that of the hybrid neural network.}} We assume the dimension of the input multiuser channel matrix  is $K\times 2N_t\times 1$ which indicates that the number of input channels (in the terminology of CNN) is $K$;  and the output dimension is $2K\times 1$ which represents the concatenated  uplink and downlink power vector.
\begin{itemize}
  \item \textbf{Classical CNN} in Fig. \ref{fig:classical_NN}. Suppose the number of filters is $F$ (so is the number of CNN output channels),  the kernel size is $m\times m$, and both padding and stride are one.  The number of parameters in the CNN module is $(Km^2+1)F$ \cite{CNN}  including $Km^2F$ weights and $F$ biases, and the batch normalization layer requires $2F$ trainable parameters. The output dimension of CNN is $F\times (2N_t-m+3) \times (1-m+3)$ and after the flatten layer, the input size to the fully connected layer is $F(2N_t-m+3)(4-m)$. Since there are $2K$ neurons in the fully connected layer, the number of associated  parameters is $2K(F(2N_t-m+3)(4-m) +1)$ where the constant 1 is due to the biases. Therefore in total the number of parameters is
      \be \label{CNN:para} P_{CNN}=(Km^2+3)F + 2K(F(2N_t-m+3)(4-m) +1).\ee
  \item \textbf{QNN} in Fig. \ref{fig:CNN_QNN}. In the proposed hybrid QNN structure, it reuses the first CNN module, so the associated number of parameters (including the batch normalization layer) is also $(Km^2+3)F$, and after flatten layer, the output dimension is $F\times (2N_t-m+3) \times (4-m)$. To generate input to the QNN with $Q$ qubits, the first fully connected layer has $Q$ neurons, so the associated number of parameters is $(F (2N_t-m+3)(4-m)+1)Q$. The QNN module with layer $L$ has $LQ$ parameters. The second fully connected layer has $2K(Q+1)$ parameters.  Therefore the total number of parameters of  this proposed hybrid structure is
      \be \scriptstyle P_{QNN}=(Km^2+3)F +(F (2N_t-m+3)(4-m)+1)Q + LQ+2K(Q+1).\ee
      Since normally a small number of qubits $Q$ is needed, the number of parameters is less than that in the classical CNN. Especially, when $K=N_t\rightarrow\infty$, we can verify that $\frac{P_{QNN}}{P_{CNN}} \rightarrow 0$, and this confirms the advantage of the proposed QNN in terms of requiring less trainable parameters.
            {{This result is significant since the number of training parameters of the proposed QNN approach only increases linearly with $N_t$ and $K$, while that of the classical CNN increases with $K N_t$. Therefore our proposed QNN has the potential to achieve more scalable performance.}}

  \item \textbf{QNN with transfer learning}  in Fig. \ref{fig:CNN_QNN}. The difference of this hybrid structure from the above one is that the weights and biases of the first CNN module is fixed and non-trainable, only the remaining layers involve trainable parameters, and therefore the number of parameters is
      \be
      P_{QNN_T}= (F (2N_t-m+3)(4-m)+1)Q + LQ+2K(Q+1),\ee which is significantly less than the classical CNN especially when the number of users is large. This will be confirmed by results in Section VI.
  \item \textbf{QCNN} in Fig. \ref{fig:QCNN}. In the second proposed hybrid structure, a  QCNN is introduced before the CNN module. The number of parameters in QCNN is $LQ$. Suppose the quantum kernel dimension is $F_Q\times F_Q$ ($F_Q=2$ in the example in Fig. \ref{fig:QCNN}). The output dimension of QCNN is $\frac{K}{F_Q}\times \frac{2N_t}{F_Q}\times Q$. Then the number of weights in  the next CNN module is
      $(\frac{K}{F_Q}m^2+3)F$, and  the output dimension of CNN is $F\times (\frac{2N_t}{F_Q}-m+3) \times (Q-m+3)$. The number of parameters in the following fully connected layer is $2K(F  (\frac{2N_t}{F_Q}-m+3)   (Q-m+3) +1)$. Therefore the total number of parameters in this hybrid structure is
      \be\label{qcNN:para} \scriptstyle
      P_{QCNN} = LQ + (\frac{K}{F_Q}m^2+3)F + 2K(F  (\frac{2N_t}{F_Q}-m+3)   (Q-m+3) +1).
      \ee
      By comparing $P_{QCNN}$ in \eqref{CNN:para} with $P_{CNN}$ in \eqref{qcNN:para}, we notice that as long as the number of qubits $Q$ is small, in general, the proposed QCNN requires less trainable parameters.  Specifically, when $K=N_t\rightarrow\infty$, we can see that $\frac{P_{QCNN}}{P_{CNN}} \rightarrow \frac{Q}{F_Q}$. Suppose the quantum kernel is a square one with dimension $F_Q\times F_Q$, then $F_Q=2^{Q/2}$ which yields $\frac{P_{QCNN}}{P_{CNN}}\rightarrow \frac{Q}{F_Q}=\frac{Q}{2^{Q/2}}$. This observation confirms the exponential quantum advantage of the proposed QNN in terms of requiring less trainable parameters.
\end{itemize}

For comparison, if the direct output of the last layer of neural networks is beamforming with dimension $2N_tK$, the numbers of trainable parameters are as shown below in Table \ref{tab:parameters}. 
\begin{table}[h]
\centering
\caption{The numbers of parameters when the direct output is beamforming.}
\label{tab:parameters}
\begin{tabular}{|p{15mm}|l|}
\hline
 Neural networks  & The number of trainable parameters \\ \hline
 \textbf{Classical CNN} &   $\scriptstyle P_{CNN}=(Km^2+3)F + 2N_t K(F(2N_t-m+3)(4-m) +1)$ \\\hline
\textbf{QNN} &   $ \scriptstyle (Km^2+3)F +(F (2N_t-m+3)(4-m)+1)Q + LQ+2 K(Q+1)$ \\\hline
\textbf{QNN with \newline transfer learning} &   $\scriptstyle  (F (2N_t-m+3)(4-m)+1)Q + LQ+2 K(Q+1)$ \\\hline
\textbf{QCNN} &   $\scriptstyle LQ + (\frac{K}{F_Q}m^2+3)F + 2N_t K(F  (\frac{2N_t}{F_Q}-m+3)   (Q-m+3) +1)$ \\\hline
\end{tabular}
\end{table}
{{By comparing Table I with the above results, it can be clearly seen that by using the power vectors as the neural network output, the trainable parameters can be significantly reduced because the output dimension is reduced by a factor of $N_t$.}}

\subsection{Training and Testing}
The training of the hybrid quantum neural network follows the same way as the classical neural network. {{We still adopt unsupervised training and the same loss function in  \eqref{loss2} and Adam optimizer in the training process}}. The training  is based on backpropagation using any classical gradient-based optimization algorithm.
When calculating the gradient on a quantum simulator without noise, the quantum circuit can be viewed as a black box and the gradient of this black box with respect to its parameters  can be calculated in a standard way.

The new challenge is to compute the gradients on a quantum computer because the noise in NISQ devices will render accurate approximation of gradients impossible. In addition, it is desirable that the same quantum circuit can be used to compute both the quantum function and the gradient of the quantum function. To achieve these goals, the parameter-shift rule \cite{gradient} is widely used to compute the gradients of quantum functions, which evaluates the original expectation in the measurement twice, but with one circuit parameter shifted by a fixed value.

{{
To be specific, suppose the loss function  is $f$, and then the derivative of the expectation value of the variational quantum circuit with respect to the parameter $\theta_i$ is computed using the parameter-shift rule  is   follows:
\be 
    \frac{d f}{d \theta_i} = r \left[f\left(\theta_i+\frac{\pi}{4r}\right)-f\left(\theta_i-\frac{\pi}{4r}\right)\right],
\ee
where $r=\frac{e_1- e_0}{2}$ is the shift constant determined by the eigenvalues $\{e_0, e_1, e_0<e_1\}$ of the Hermitian generator of the unitary gate parameterized by $\theta_i$.
}}

It is similar to the standard gradient derivation that uses two evaluations of the a function at proximity. However, in the parameter-shift rule, the shift between two data points is finite and is not infinitesimal.  The parameter-shift rule is supported in most existing quantum simulators such as Pennylane \cite{pennylane}, Qiskit  \cite{qiskit}, TensorFlow Quantum \cite{TensorFlow-Quantum} and Amazon Braket \cite{Braket} simulators. The testing phase follows a standard approach as the classical neural network and the details are omitted.

\section{Quantum Noise}
So far we have assumed the quantum operations are perfect but quantum computing is still a nascent technology, and quantum noise is the major limiting factor of the NISQ devices. This section will present some example models of quantum noise \cite{Nielsen-book} that will be used to evaluate the performance of our proposed hybrid neural networks in the next section.

A bit flip error is a type of error in which  the state of a qubit is flipped from $\ket{0}$ to $\ket{1}$ or vice versa with probability $p$. This could happen during hardware measurement or calibration. Quantum noise is commonly modelled by quantum channels that is represented by Kraus operators $\{\qK_i\}$ satisfying the condition $\sum_i \qK_i^\dag \qK_i= \qI$. Assume the initial status of a qubit is $\rho$, then output status after passing a channel is written as:
\be
    \Phi(\rho) = \sum_i \qK_i^\dag \rho \qK_i.
\ee
The effect of a channel is to apply a  random unitary   transformation $\qK_i$ with a certain probability. The Kraus operators of a bit flip channel can be expressed as:
\be
    \qK_0 = \sqrt{1-p} \left[\begin{matrix}
1 & 0 \\
 0 & 1
\end{matrix}\right], \qK_1 = \sqrt{p}X = \sqrt{p} \left[\begin{matrix}
0 & 1 \\
 1& 0
\end{matrix}\right].
\ee
Note that flipping the state of a qubit is equivalent to applying an X gate.

Similarly we can define the phase flip channel, whose Kraus operators are
\be
    \qK_0 = \sqrt{1-p} \left[\begin{matrix}
1 & 0 \\
 0 & 1
\end{matrix}\right], \qK_1 = \sqrt{p}Z = \sqrt{p} \left[\begin{matrix}
1 & 0 \\
 0& -1
\end{matrix}\right].
\ee
In other words,  the phase flip channel applies a Z gate with the probability $p$.

The depolarizing channel is a generalization of the bit flip and phase flip channels, in which each of the three types of Pauli errors happen with the same probability. Its Kraus operators are given by
{\bea
    \qK_0 = \sqrt{1-p} \left[\begin{matrix}
1 & 0 \\
 0 & 1
\end{matrix}\right], \qK_1 = \sqrt{\frac{p}{3}}X = \sqrt{\frac{p}{3}}   \left[\begin{matrix}
0 & 1 \\
 1& 0
\end{matrix}\right],\\
 \qK_2 =  \sqrt{\frac{p}{3}}Y =  \sqrt{\frac{p}{3}} \left[\begin{matrix}
0 & -i \\
 i& 0
\end{matrix}\right],   \qK_3 =  \sqrt{\frac{p}{3}}Z =  \sqrt{\frac{p}{3}} \left[\begin{matrix}
    1 & 0 \\
     0& -1
    \end{matrix}\right].
\eea}
We can see the depolarizing channel leaves the status of a qubit unchanged with the probability $1-p$, and apply X, Y and Z gates each with the equal probability $\frac{p}{3}$. In the next section, we will show the effect of the bit flip channel and the depolarizing channel using simulation results.

\section{Simulation Results}\label{simu}
In this section, numerical simulations are carried out to evaluate the proposed hybrid neural networks for optimizing the  downlink beamforming in a wireless communications network.  We consider an MISO system in which one BS serves multiple users using 20 MHz bandwidth and the noise power spectral density is -174 dBm/Hz.
The large-scale fading is considered and the path loss is modelled as $128.1 + 37.6\log_{10}(d)$ where $d$ is the distance in km between the BS and a user. We assume users are uniformly located in an area between 100 and 500 meters from the BS. The small-scale fading follows the Rayleigh distribution with zero mean and unit variance. We assume the number of antennas at the BS is the same as the number of users, i.e., $N_t=K$. We employ the deep learning software  Pytorch and the PennyLane framework  to implement the proposed quantum neural networks. Unless otherwise specified, we use 10,000  channel samples for training the neural networks with 150 epochs, a minibatch size of 100 and a learning rate of 0.001 of which 20\% of data samples are used for validation; and 1,000 channel samples are used for testing the performance of the trained neural networks. The transmit power is assumed to be 30 dBm.

{{For comparison, we consider the classical neural network (NN) introduced in Section II as a benchmark, and the WMMSE solution  in \cite{shi2011iteratively}  which is obtained by using a large number of iterations to serve as a performance upper bound. The sum rate is normalized as a percentage by the WMMSE solution for all other schemes so it is always less than one.}}

First we will decide the QNN related hyper parameters including the number of layers and number of qubits by studying the performance of the proposed  hybrid structures for a system with $N_t=K=8$. We vary the number of quantum circuit layers from 2 to 6 while fixing the number of qubits to be $Q=4$. The results are shown in Table \ref{tab:layer}. It is observed that  increasing the number of layers does not have an obvious effect on the resulting sum rate. Next in Table \ref{tab:qubit} we show the results by varying the number of qubits from 2 to 8 while keeping the number of layers to be $L=2$ and again, there is no significant effect on the achievable sum rate. Therefore, in the sequel we choose $L=2$ layers and $Q=4$ qubits for low complexity unless otherwise specified.

\begin{table}[h]
\centering
\caption{The effect of the number of quantum circuit layers, $Q=4$.}
\label{tab:layer}
\begin{tabular}{|l|l|l|l|l|}
\hline
 $L$ & 2  & 3  & 4  & 6 \\ \hline
 $R_{QNN}$ &   0.9202   & 0.9168  & 0.9232   & \textbf{0.9259} \\\hline
\end{tabular}
\end{table}

\begin{table}[h]
\centering
\caption{ The effect of the number of qubits in the designed quantum circuit, $L=2$.}
\label{tab:qubit}
\begin{tabular}{|l|l|l|l|l|}
\hline
 $Q$ & 2  & 4  & 6  & 8 \\ \hline
 $R_{QNN}$ &   0.9165  & \textbf{0.9202}  &0.9201  & 0.9168\\\hline
\end{tabular}
\end{table}

Next  we show the evolution of the loss function for different neural networks in both training and validation processes in Fig. \ref{fig:losses}  for a system with  $N_t = K = 8$. First note that for all schemes, the values of loss functions keep decreasing until convergence and this confirms the feasibility and trainability of the proposed quantum neural networks. {{For the classical NN, its training loss first reduces quickly but then takes longer time to converge, and it has the largest gap between the training and validation losses among all schemes. This may indicate that the learned model is underfitting and cannot accurately capture the relationship between the input channel and output beamforming.  For the proposed  QNN method, it shows slightly faster convergence than the classical NN and its training and validation losses match well with each other. We also observe that for QNN, the validation losses are lower than the training loss at first and then become similar and slighter higher when the training finishes. This is  because the training loss is calculated during each epoch, but validation loss is calculated at the end of each epoch and by then the neural network model is much better trained and improves significantly. The QNN method with transfer learning shows faster convergence than QNN and classical NN because of pretraining, but their training and validations losses are high. The proposed QCNN  method  achieves the fastest convergence but we also notice that its training loss remains high. This indicates that it may have experienced the barren plateau phenomenon in training and how to overcome this is worth further study. In addition, note that in general QCNN introduces more floating-point operations (FLOPs) which could delay   the training process.}}
\begin{figure}[t]
\centering
\includegraphics[width=3in]{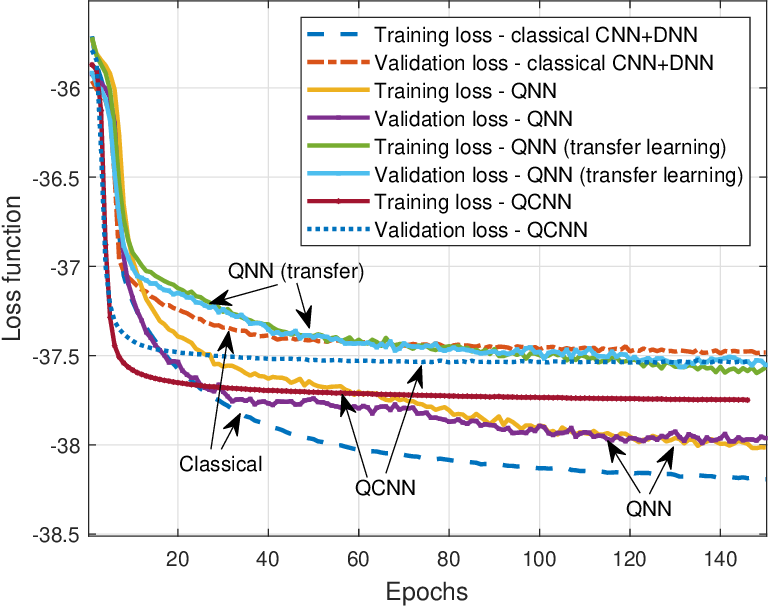}
\caption{ The training and validation losses vs the number of epochs when $N_t =K=8$.}
\label{fig:losses}
\end{figure}

Then we compare the sum rate performance for different number of users (and BS antennas) in Table \ref{tab:sumrate:K}. As can be seen, both the proposed QNN and transfer learning methods achieve similar sum rate performance as the classical NN. The proposed QCNN method achieves higher data rates especially as the number of users increases, and this confirms that the QCNN can effectively extract more useful features than classical CNN. This performance gain is important considering that only two qubits are used by QCNN due to the use of amplitude embedding.
The effect of the embedding methods is evaluated in Table \ref{tab:embedding}, and for convenience of comparison the results in Table \ref{tab:sumrate:K} are also included. We can see that for the proposed QNN method including transfer learning, using either the angle embedding or the amplitude embedding does not affect much the sum rate. However, for the proposed QCNN method, there is a significant performance loss with the angle embedding. To reiterate, we have adopted angle embedding for QNN and amplitude embedding for QCNN.  We observe that while the performance of both CNN and QNN degrade as the number of users increases,  QCNN can maintain a more stable performance and this verifies its advantage in terms of scalability.
\begin{table}[h]
\centering{
\caption{ The sum rate performance vs. the number of users }
\label{tab:sumrate:K}
\begin{tabular}{|l|l|l|l|l|l|}
\hline
 $K=N_t$    & 4  & 6  & 8 & 10 & 20\\ \hline
 $R_{Classical}$ &  0.9330  & 0.9134 &0.9135 & 0.9103 &0.8988\\ \hline
 $R_{QNN}$ & 0.9320   & 0.9235  & \textbf{0.9191}  & 0.9088  & 0.9023 \\ \hline
  $R_{QNN (Transfer)}$ & \textbf{0.9346}   & 0.9114  & 0.9112 & 0.9151  & 0.9094\\ \hline
 $R_{QCNN}$ &  0.9256  & \textbf{0.9262} &\textbf{0.9149}  & \textbf{0.9241}  &\textbf{0.9223}\\ \hline
\end{tabular}}
\end{table}

\begin{table}[h]
{\tiny
\centering
\caption{ The effect of the embedding methods on the sum rate performance}
\label{tab:embedding}
\begin{tabular}{|l|l|l|l|l|l|}
\hline
 $K=N_t$    & 4  & 6  & 8 & 10 & 20\\ \hline
 $R_{QNN}, angle$ & \textbf{0.9320}   & 0.9235  & 0.9191  & 0.9088  & 0.9023 \\ \hline
 $R_{QNN}, amplitude$ & 0.9204   & 0.9226  & 0.9097  & 0.8958  & 0.8938 \\ \hline
  $R_{QNN (Transfer)}, angle$ & 0.9346   & 0.9114  & 0.9112 & 0.9151  & 0.9094\\ \hline
  $R_{QNN (Transfer)}, amplitude$ & 0.9348   & 0.9224   &0.9155   & 0.8910   & 0.9013 \\ \hline
 $R_{QCNN}, amplitude$ &  0.9241 & \textbf{0.9238} &\textbf{0.9164}  & \textbf{0.9215}  &\textbf{0.9275}\\ \hline
  $R_{QCNN}, angle$ & 0.9203    & 0.8841  &0.8698   & 0.8931   &0.8766\\ \hline
\end{tabular}}
\end{table}

In addition to the sum rate, we also investigate the number of trainable parameters in the proposed quantum neural networks in Fig. \ref{fig:parameter} {{assuming equal number of users and antennas, i.e., $K=N_t$}}. We still use the classical NN as the benchmark, so the results in Fig. \ref{fig:parameter} are normalized by the amount of parameters in the classical NN. We assume that in the CNN module used in  the classical NN and the QCNN, the number of filters is $F=8$, and kernel dimension is $m=3$. In the proposed QCNN, the kernel is of size $2\times 2$ as shown in Fig. \ref{fig:QCNN} and the number of qubits is thus two.
It is observed that both QNN methods require significantly less parameters than the classical NN. For instance,  the QNN method and the QNN with transfer learning require only about 20\% and 10\% parameters, respectively, for a 20-user system.
This advantage is even more obvious when the number of users increases. {{This is expected as we can see from  the analysis in Section III.C, for the classical NN, when $N_t=K$ the number of trainable parameters increases with $K^2$ while that of the QNN increases linearly with $K$.   The QCNN method requires more parameters than QNN but in general they are still less than the classical NN as detailed in the analysis after \eqref{qcNN:para}}}, and QCNN has superior sum rate performance as shown in Table \ref{tab:sumrate:K}. By combining the results in Fig. \ref{fig:parameter} and Table \ref{tab:sumrate:K}, we can clearly verify the advantages of the proposed hybrid quantum-classical neural networks, i.e., they can either improve the sum rate performance or reduce the amount of trainable parameters required when compared to the classical NN.

\begin{figure}[h]
\centering
\includegraphics[width=3in]{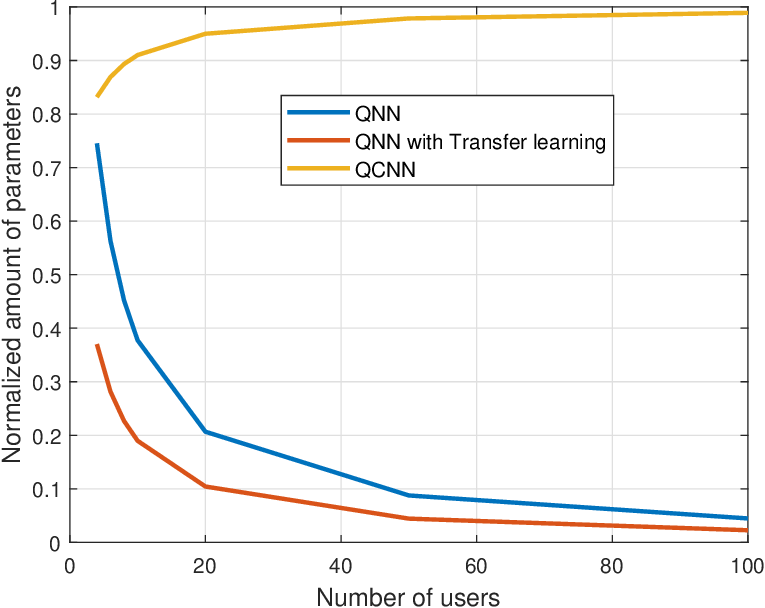}
\caption{ The normalized amount of trainable parameters in the proposed quantum neural networks with hyper parameters $F=8, m=3, L=2, Q=4 ~~\mbox{(QNN)}$ or $Q=2 ~~\mbox{(QCNN)}$.}
\label{fig:parameter}
\end{figure}

{{It is obvious that the quality of channel state information is important for the optimization of beamforming, so next we evaluate the impact of imperfect channel estimation on the proposed QNN method. We assume that the real channel between the $j$-th  antenna and the $k$-th user is expressed as  $h_{j,k} = \hat h_{j,k} + \Delta h_{j,k}$, where $\hat h_{j,k}$ is the available channel estimate for neural network training and inference while $\Delta h_{j,k}\in  \mathcal{CN}(0,\sigma^2_e)$ denotes the channel estimation error with variance of $\sigma^2_e$. In Table \ref{tab:CSI}, we show the results of the achievable rate when $\sigma^2_e$ varies from 0.01 to 0.2 for both the classical neural network and the proposed QNN method. It can be observed that the rates for both methods decrease  gracefully as the channel estimation error increases and the proposed QNN method still outperforms the classical neural network.}}
\begin{table}[h]{
\centering
\caption{ The effect of the channel estimation error, $K=N_t=8$.}
\label{tab:CSI}
\begin{tabular}{|l|l|l|l|l|l|}
\hline
 $\sigma^2_e$ & 0.01  & 0.05  & 0.1  & 0.15 & 0.2\\ \hline
  $R_{Classical}$ &  0.9071  & 0.9008  &0.8931  & 0.8859& 0.8789\\\hline
 $R_{QNN}$ &   0.9202  & 0.9139  &0.9062  & 0.8988& 0.8917\\\hline
\end{tabular}}
\end{table}

So far we have assumed ideal quantum circuit operation using Pennylane's `default.qubit' device. However, current NISQ devices  cannot execute quantum circuits perfectly, so next we will investigate the performance in noisy environments. Note that due to the excessively long time needed to simulate the noisy circuit using IBM's simulators/devices, we only consider the noise in the test phase to speed up the simulation while in the training phase, we assume the circuits are perfect. Consequently, the results presented here are a lower bound of the actual performance but our simulation results for small systems verify that the performance is close to the scenario when noise is also considered in the training phase. Firstly, to model the error, we  simulate two noisy channels commonly used to model experimental imperfections introduced in Section V.  The first example is the simple bit flip channel, which swaps $\ket{0}$ and $\ket{1}$ with the probability $p$; and the second example is the depolarizing channel in which each of the three possible Pauli errors can be applied to a single qubit with each probability $\frac{p}{3}$. The results are shown in Table \ref{tab:noise} as the number of users varies. We also simulate the error performance by varying the error probability from 0.01 to 0.5 for a 8-user system and results are shown in Table \ref{tab:noise2}. By comparing the results in Table \ref{tab:sumrate:K} and Table \ref{tab:noise} and  \ref{tab:noise2},  we observe that as the number of users increases or the error probability increases both types of  noise cause significant performance degradation to the sum rate.  
Note that we have shown the effects of very high error probabilities, but in realistic quantum hardware, the error probability is much lower and its effect will be studied below.  
\begin{table}[h]
\centering{
\caption{ The effect of the quantum circuit noise.}
\label{tab:noise}
\begin{tabular}{|l|l|l|l|l|}
\hline
 $K$ & 4  & 6  & 8  & 10 \\ \hline
 Depolarization, $p=0.05 $ &   0.9313   & 0.9217  & 0.9174 & 0.9065 \\\hline
 Bit flip, $p=0.05$ &  0.9284   &  0.9116  & 0.9123  & 0.9007 \\\hline
  Depolarization, $p=0.3$ &   0.9213   & 0.8863  & 0.8783  & 0.8682 \\\hline
 Bit flip, $p=0.3$ & 0.9196    & 0.8846   & 0.8672 & 0.8601 \\\hline
\end{tabular}}
\end{table}

\begin{table}[h]
\centering{
\caption{ The effect of the quantum circuit noise probability, $K=8$.}
\label{tab:noise2}
\begin{tabular}{|l|l|l|l|l|l|l|}
\hline
 $p$ & 0.01 & 0.1  & 0.2  & 0.3  & 0.4 & 0.5\\ \hline
 Polarization  & 0.9189  &   0.9123  &  0.8941 & 0.8779 & 0.8710 & 0.8674 \\\hline
 Bit flip  & 0.9187 &  0.8937  &  0.8714  & 0.8674 & 0.8667 & 0.8658 \\\hline 
\end{tabular}}
\end{table}

Secondly, to assess the realistic effect of the noise without running the proposed algorithms on real quantum computers which is time-consuming, we resort to IBM quantum hardware emulators.  IBM regularly calibrates the quantum hardware available in their cloud. Its quantum software development kit Qiskit provides various hardware emulators that use the calibration data to replicate the target hardware output.  In this experiment, we use three quantum emulators, i.e., FakeLima, FakeVigo and FakeCasablanca with different numbers of qubits. These quantum systems are also different in terms of coupling map, basis gates, qubit properties (T1, T2, error rate, etc.) which are useful for performing noisy simulations of the system \cite{ibmfake}. From the results in Table \ref{tab:emulator}, we can see that our proposed methods are robust to quantum  errors on a variety of quantum emulators.

 \begin{table}[h]
\centering
\caption{The sum rate performance on IBM quantum hardware emulators.}
\label{tab:emulator}
\begin{tabular}{|l|l|l|l|}
\hline
 $K=8$ &\pbox{2.2cm}{FakeLima  \\ (5 qubits)}      & \pbox{2.2cm}{FakeVigo\\(5 qubits)}  & \pbox{2.2cm}{FakeCasablanca \\(7 qubits)}   \\ \hline
  $Q=4$ &   0.9191  & 0.9189   & 0.9189     \\\hline
  $Q=6$ &  0.9144     &0.9144   & 0.9142   \\\hline
\end{tabular}
\end{table}

To better interpret the above error results of emulators and make meaningful comparison with the bit flip noise model, we plot the mean square error (MSE) of the QNN output (after measurement before the next fully connected layer) in Fig. \ref{fig_QNN_out_err} when considering the bit flip noise and the hardware emulators. {{As can be seen, when the bit flip noise probability increases, the MSE performance degrades quickly which leads to degraded sum rate performance as verified in Table \ref{tab:noise2}.
We then examine the output MSEs of the three emulators FakeLima, FakeVigo and FakeCasablanca to infer their  bit flip noise probabilities. It is observed that all three emulators have low output MSEs   which correspond to low bit flip noise probability of 0.01 for FakeVigo and even less for the other two emulators, and this explains why their sum rate performance does not deviate much from the ideal case.}}
\begin{figure}
\centering
 \includegraphics[width=3in]{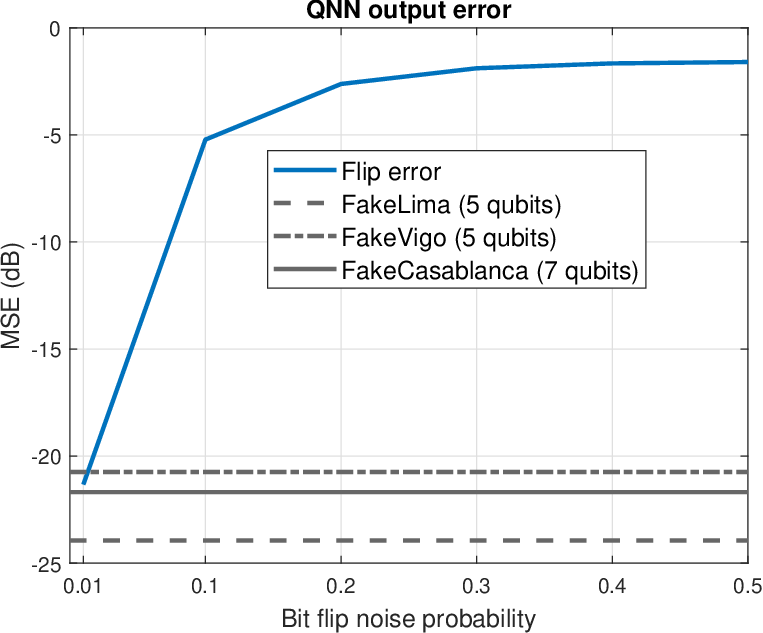}
 \caption{ MSE of the QNN output with bit flip error and emulators, $Q=4$.}
 \label{fig_QNN_out_err}
\end{figure}

\section{Conclusions}\label{conc}
 In this paper, we introduced quantum machine learning to enhance the performance of classical neural networks to optimize multiuser beamforming, leading to a hybrid quantum-classical neural network framework. We proposed two specific methods to combine quantum neural network and classical neural network.  Simulation results demonstrated that the proposed methods are capable of solving the multiuser beamforming problem, and can achieve advantages over classical neural networks in terms of improved performance or reduced amount of trainable parameters. Through noise models and hardware emulators, the proposed algorithms are also shown to be robust on NISQ devices for small errors.

 The proposed method shows that tools from quantum machine learning can be used to enhance classical solvers for the multiuser beamforming problem. Performance advantages of such enhanced solvers manifest already with relatively small sizes of the quantum system.  The proposed hybrid quantum-classical neural networks are also useful for other wireless-related purposes, particularly concerning the projected growth of user terminals such as  non-orthogonal multiple access, power control in extremely large MIMO systems, Internet of Things networks, federated learning in mobile edge computing and integrated satellite-terrestrial networks.

This observation necessitates a more careful investigation of the problem in several important directions including but not limited to the identification of potential bottlenecks arising in implementations, quantum-inspired solvers for solvers which use only classical neural networks, channel variations caused by user mobility, error mitigation techniques to minimize the effects of noise and architectural design questions.

 In particular, the practical implementation of quantum neural networks still faces major challenges. The energy consumption of quantum computers is still high and cannot be afforded by a single base station. The latency budgets of the overall architecture will need to be investigated. For this, the 5G centralized radio access network (C-RAN) architecture could be a potentially solution to implement quantum computing with low latency \cite{Annealing-MIMO}\cite{QA-uplink}\cite{quantum-6G}.

\section*{Appendix: Quantum Preliminaries}\label{Preliminaries}
This section provides background and fundamentals on quantum computing    necessary to the development of our proposed quantum neural networks to solve the sum rate maximization problem  \eqref{generalproblem}. For a comprehensive coverage on quantum computation and quantum information, readers are referred to textbooks or surveys such as \cite{Nielsen-book} and \cite{quantum-search}.

\subsection{Qubits}
Analogous to  bits used by classical computing as the smallest unit, quantum computing uses qubits. Unlike a classical bit, a qubit can be in a superposition state at the same time. The state of a qubit can be expressed using the orthogonal computational basis of   $\{\ket{0}, \ket{1}\}$ in a two-dimensional Hilbert space as
\be\label{singlequbit}
    \ket{\psi} = a \ket{0} + b\ket{1},
\ee
{{where $a$ and $b$ are  probability amplitudes that are complex coefficients and satisfy $|a|^2+ |b|^2=1$}}. The Dirac's ``bracket'' notation $|\cdot\rangle$ is called a ket which is used to indicate that the object is a column vector. The complex conjugate transpose of $|\cdot\rangle$ is denoted as $\langle \cdot|$, which is called a bra. The state of a qubit can also be equivalently written as a vector whose elements are the probability amplitudes.

The power of quantum computing lies in the interaction of multiple qubits. Consider a two-qubit system in which the two qubits states are expressed as  $\ket{\psi_1}= a_1 \ket{0} + b_1\ket{1}$ and $\ket{\psi_2} = a_2 \ket{0} + b_2\ket{1}$. The state $|\psi\rangle$ of the joint system is then written as   the tensor product of the quantum states of the two qubits, i.e.,
{
\bea\label{multiqubits}
    \ket{\psi}&=& \ket{\psi_1\psi_2}= \ket{\psi_1}\otimes \ket{\psi_2} \notag\\
    &=& a_1 a_2 \ket{00} + a_1b_2\ket{01} + a_2b_1\ket{10} + a_2b_2\ket{11}.
\eea}
 In general, an $n-$qubit system has $2^n$ probability amplitudes each corresponding to an orthogonal basis.

A quantum state can also be characterized by its state vector  with the length being the number of orthogonal states and each elements representing the corresponding amplitude. For instance, for the single qubit  in \eqref{singlequbit}, its state vector is
\be\label{singlequbit:vector}
    \ket{\psi} =  [a~~ b]^T,
\ee
while for the state of the two qubits in \eqref{multiqubits}, its state vector is
\be\label{multilequbit:vector}
    \ket{\psi} =  [a_1 a_2 ~~a_1b_2~~ a_2b_1~~a_2b_2]^T.
\ee

In the above example, $\ket{\psi_1}$ and $\ket{\psi_2}$ are independent of each other.  There also exists scenarios in which
the quantum states of multiple qubits are correlated and cannot be represented separately and independently of individual qubit states. In this case, the qubits are entangled with each other. Bell states  are the simplest example of quantum entanglement of two qubits and include the following four maximally entangled states of two qubits:
{\bea\label{bell}
    \ket{\Phi^+} &=& \frac{1}{\sqrt{2}} \ket{00} + \frac{1}{\sqrt{2}} \ket{11} \notag \\
    \ket{\Phi^-} &=& \frac{1}{\sqrt{2}} \ket{00} - \frac{1}{\sqrt{2}} \ket{11} \notag \\
    \ket{\Psi^+} &=& \frac{1}{\sqrt{2}} \ket{01} + \frac{1}{\sqrt{2}} \ket{10} \notag \\
    \ket{\Psi^-} &=& \frac{1}{\sqrt{2}} \ket{01} - \frac{1}{\sqrt{2}} \ket{10}.
\eea}
It can be verified that Bell states in \eqref{bell} cannot be described as the form in \eqref{multiqubits}.  Quantum entanglement has important implications and this property allows quantum computers to perform operations that are difficult to achieve on  classical
computers.  The existence of entanglement has been experimentally verified by the CHSH game \cite{CHSH}, which is a thought experiment involving two cooperating  players who want to win a game without communicating with each other during the game.  It is shown that if
both players use only classical strategies involving their local information and potentially shared randomness, the maximum probability they win is 75\%. However, if both players pre-share an entangled qubit pair, e.g., $\ket{\Phi^+}$, then using a quantum strategy they succeed with a probability of ~85\%. This experiment gives strong evidence of quantum entanglement.

\subsection{Measurement}
Measurements extract information about a quantum system and link it to a classical system. For a single qubit state in \eqref{singlequbit}, the complex amplitudes $a$ and $b$ determine the probabilities of obtaining $\ket{0}$ or $\ket{1}$ when {{measuring the state of a qubit  relative to a given computational  basis. Suppose we use the Z basis of   $\{\ket{0}, \ket{1}\}$ with corresponding eigenvalues $\lambda_0=1, \lambda_1=-1$, then the probability of obtaining the outcome $j$ is
\be
    p(j) = \bra{\psi}P_j P_j^\dag \ket{\psi} = \left\{ \begin{matrix}
|a|^2, & P_j =\ket{0};\\
 |b|^2, & P_j =\ket{1}.
\end{matrix} \right.
\ee
That is, after  measurement,  the probabilities of obtaining  $\ket{0}$  and $\ket{1}$ are   $|a|^2$ and  $|b|^2$, respectively.
The expected value after the measurement is then calculated as $\sum_{j=0,1} p(j)\lambda_j$.

{{Note that once an outcome is observed after the measurement, the quantum state collapses to a classical bit \footnote{{{There also exists non-destructive measurement that can infer how the measurement process changes the quantum state of the system.}}}, and all quantum operations will have the same effect as classical operations. In other words, the quantum computation ends with the qubit measurement.}}  In practice,  measurements are often restricted to the Z-basis such as on IBM quantum computers, so each qubit needs to be first rotated to the Z-basis before measurement.

\subsection{Quantum Gates}
Quantum gates are basic quantum circuits that can be used to change the quantum state of a single or multiple qubits in a reversible way.
Quantum gates are unitary operators and can be interpreted as rotations of the quantum state vector in the Bloch sphere. Applying a quantum gate to qubit(s) is equivalent to left multiplying the corresponding unitary matrix with the state vector representation of qubits. Below we describe some useful quantum gates that are necessary in our proposed quantum neural networks.

\begin{itemize}
\item We start from the Hadamard gates for a single qubit.
The Hadamard gate   creates equiprobable  superpositions of the two states and can be represented by the matrix
\be
 H = \frac{1}{\sqrt{2}}\left[\begin{matrix}
1 & 1 \\
 1& -1
\end{matrix}\right],
\ee
and it performs the transforms below
{\bea
     H \ket{0} = \frac{1}{\sqrt{2}} \ket{0} + \frac{1}{\sqrt{2}}\ket{1},\notag\\
    H \ket{1} = \frac{1}{\sqrt{2}} \ket{0} - \frac{1}{\sqrt{2}}\ket{1}.
\eea}

\item Another set of three popular single-qubit quantum gates belonging to the Pauli gates are expressed as
\be
    X = \left[\begin{matrix}
0 & 1 \\
 1& 0
\end{matrix}\right],
    Y = \left[\begin{matrix}
0 & -i \\
 i& 0
\end{matrix}\right],
    Z = \left[\begin{matrix}
1 & 0 \\
 0& -1
\end{matrix}\right].
\ee
We can see that the X gate swaps the amplitudes of the two states, i.e.,
\be
X \ket{\psi} =b \ket{0} + a\ket{1},
\ee
so it is also known as the NOT gate. Y gate and Z gate perform rotations by $\pi$  around the $y$-axis and $z$-axis of the Bloch sphere, respectively, which can be verified as:
\be
Y \ket{\psi} = -i(b \ket{0} + a\ket{1}), Z \ket{\psi} = a \ket{0} -b \ket{1}.
\ee

\item Different from the above gates with pre-determined operations, the $R_y(\theta)$ gate is a parameterized gate that performs the rotation of a qubit about the Y-axis in the Bloch sphere by a given parameter angle $\theta$.  $R_y(\theta)$ gate can be expressed as
\be
R_y(\theta)=\left[\begin{matrix}
\cos(\frac{\theta}{2}) & -\sin(\frac{\theta}{2}) \\
 \sin(\frac{\theta}{2})& \cos(\frac{\theta}{2})
\end{matrix}\right].
\ee

\item   The Controlled-NOT (CNOT) gate is a popular two-qubit gate, which is used  to entangle two qubits  and is essential in quantum computing. This gate is a conditional gate that applies an X-gate (NOT operation) on the second qubit, if the state of the first qubit  is $\ket{1}$. The first and the second qubits are often called the control and target qubits, respectively. Mathematically, a CNOT gate for two qubits can be written as
\be
\mbox{CNOT} = \left[ \begin{array}{cccc}
                1 & 0 & 0 & 0 \\
                0 & 1 & 0 & 0 \\
                0 & 0 & 0 & 1 \\
                0 & 0 & 1 & 0 \\
             \end{array}\right].
\ee
It is easy to verify that if we apply the CNOT gate to the $\ket{10}$ state which is represented as the state vector $[0~~ 0~~ 1~~ 0]^T$,  we will get the result of $[0 ~~ 0 ~~ 0~~ 1]^T$ which corresponds to the state of $\ket{11}$. This is because  the control qubit is $\ket{1}$,  the target qubit needs to be flipped from $\ket{0}$ to $\ket{1}$.  
\end{itemize}

 Note that qubits can be realized through various physical systems like superconducting circuits, trapped ions, or photonic systems. Take the superconducting quantum  computers for example. They use Josephson junctions to build artificial atoms as superconducting qubits and by irradiating microwave photons at these superconducting qubits, one can control their behavior including the above mentioned gate operations \cite{ibmqc}. Superconducting quantum computers have the fastest gate speed which is measured in nanoseconds, so it is sufficient to support the computational tasks  for 5G/6G communications.

\end{document}